\documentclass[11pt]{article}

\usepackage[utf8]{inputenc}
\usepackage{lipsum}

\usepackage{amsmath, amssymb,amsfonts,amsthm,mathtools}
\usepackage{csquotes}
\usepackage{xspace,graphicx,relsize,bm,xcolor}
\usepackage{soul} % provides  s p a c i n g o u t, underlining and some derivatives such as overstriking and highlighting: https://ctan.mc1.root.project-creative.net/macros/generic/soul/soul.pdf
\usepackage{nicefrac}

\usepackage{parskip}  % no indentation but some space between paragraphs

\usepackage{libertine}
\usepackage{libertinust1math}
\usepackage{dsfont}
\usepackage[T1]{fontenc}
\usepackage{enumitem}

\usepackage[
    backend=biber,
    style=alphabetic,
    sorting=anyt,
    minalphanames=3,
    maxalphanames=3,
    maxnames=99,
    backref=true
    ]{biblatex}
\DefineBibliographyStrings{english}{%
  backrefpage = {page},% originally "cited on page"
  backrefpages = {pages},% originally "cited on pages"
}

\usepackage{sepfootnotes}
\newendnotes{x}
\renewcommand\xnotesize\normalsize

\usepackage{hyperref}
\usepackage[margin=1.75cm]{geometry}
\definecolor{linkcol}{rgb}{0.0,0.55,0.7}
\definecolor{citecol}{rgb}{0.0, 0.6, 0.45}
\definecolor{urlcol}{rgb}{0.7, 0.0, 0.55}
\hypersetup{
	colorlinks,
	linkcolor={linkcol},
	citecolor={citecol},
	urlcolor={urlcol}
}

\usepackage{url}
\usepackage{subcaption}
\usepackage{mleftright}
\usepackage{hyperref}
\usepackage{multirow}
\usepackage{physics}  % all braket notation at once

\usepackage{algorithm}
\usepackage{algpseudocode}

\usepackage{cleveref}

\usepackage{authblk}  % for Authors

\def\01{\{0,1\}}

\newtheoremstyle{mydefinitionsty}% 〈name〉
{10pt}% 〈Space above〉
{10pt}% 〈Space below〉
{}% 〈Body font〉
{}% 〈Indent amount〉1
{}% 〈Theorem head font〉
{}% 〈Punctuation after theorem head〉
{.5em}% 〈Space after theorem head〉2
{\textbf{\thmname{#1}~\thmnumber{#2}:  }\thmnote{(#3)}}% 〈Theorem head spec (can be left empty, meaning ‘normal’)〉
\theoremstyle{mydefinitionsty}

\newtheorem{observation}{Observation}

\newtheoremstyle{myproblemsty}% 〈name〉
{10pt}% 〈Space above〉
{10pt}% 〈Space below〉
{}% 〈Body font〉
{}% 〈Indent amount〉1
{}% 〈Theorem head font〉
{}% 〈Punctuation after theorem head〉
{.5em}% 〈Space after theorem head〉2
{\textbf{\thmname{#1}~\thmnumber{#2}:  }\thmnote{(#3)}\newline}% 〈Theorem head spec (can be left empty, meaning ‘normal’)〉
\theoremstyle{myproblemsty}

\newtheoremstyle{mythmsty}% 〈name〉
{10pt}% 〈Space above〉
{10pt}% 〈Space below〉
{\itshape}% 〈Body font〉
{}% 〈Indent amount〉1
{}% 〈Theorem head font〉
{}% 〈Punctuation after theorem head〉
{.5em}% 〈Space after theorem head〉2
{\textbf{\thmname{#1}~\thmnumber{#2}:  }\thmnote{(#3)}}% 〈Theorem head spec (can be left empty, meaning ‘normal’)〉
\theoremstyle{mythmsty}

\definecolor{alexcolor}{rgb}{0.0, 0.47, 0.75}   % {0.27, 0.51, 0.71}  %

\definecolor{questioncolor}{rgb}{0.36, 0.54, 0.66}

\title{Kernel-based dequantization of variational QML without Random Fourier Features}

\author[1,2,3]{Ryan Sweke}
\author[4,5]{Seongwook Shin}
\author[4,6]{Elies Gil-Fuster}

\affil[1]{African Institute for Mathematical Sciences, South Africa}
\affil[2]{Department of Mathematical Sciences, Stellenbosch University, Stellenbosch 7600, South Africa}
\affil[3]{National Institute for Theoretical and Computational Sciences (NITheCS), South Africa}
\affil[4]{Dahlem Center for Complex Quantum Systems, Freie Universität Berlin, Berlin, Germany}
\affil[5]{Department of Physics and Astronomy, Seoul National University, 08826 Seoul, South Korea}
\affil[6]{Fraunhofer Heinrich Hertz Institute, 10587 Berlin, Germany}

\date{\today}

\begin{document}
\maketitle

\begin{abstract}
\noindent There is currently a huge effort to understand the potential and limitations of variational quantum machine learning (QML) based on the optimization of parameterized quantum circuits. 
Recent proposals toward \emph{dequantizing} variational QML models for regression problems include approaches based on kernel methods with carefully chosen kernel functions, approximated via Random Fourier Features (RFF). In this short note we show that for a wide range of instances, this approach can be simplified. More specifically, we show that the kernels whose evaluation is approximated by RFF in this dequantization algorithm can in fact often be evaluated exactly and efficiently classically. As such, our results enhance the toolkit for kernel-based dequantization of variational QML.
\end{abstract}

\section{Introduction}\label{s:intro}

Variational quantum machine learning (QML) based on the optimization of parameterized quantum circuits (PQCs) is currently the subject of significant theoretical and experimental interest~\cite{benedetti2019parameterized, cerezo2021variational,schuld2018supervised}.
In particular, there is a huge effort to understand both the potential and limitations of this approach to QML, and more specifically, whether or not there exist any practically relevant settings in which this approach offers any advantages over classical learning algorithms. To this end, significant effort has been spent towards showing when and how variational QML models are \textit{not} expected to be advantageous -- typically by developing efficient classical algorithms which can be shown to achieve similar performance to variational QML for certain tasks of interest~\cite{angrisani2025simulatingquantumcircuitsarbitrary,lerch2024efficientquantumenhancedclassicalsimulation,bermejo2024quantumconvolutionalneuralnetworks,cerezo2024doesprovableabsencebarren,landman2022classicallyapproximatingvariationalquantum,Sweke2023potential,shin2024,Schreiber_2023}. Often, this approach is referred to as \textit{dequantization} of variational QML~\cite{gilfuster2025relation}. 

Along these lines, a variety of recent works have proposed and explored the use of well-designed classical kernel methods for the dequantization of PQC optimization for regression problems~\cite{landman2022classicallyapproximatingvariationalquantum,Sweke2023potential,shin2024}.
At a high-level, all of these works start from the fundamental observation that, given a specific PQC there exists a family of kernels, which can be explicitly specified, whose reproducing kernel Hilbert space contains all functions expressible by the PQC. As such, a natural strategy for dequantizing regression via optimization of this specific PQC is to choose one such kernel, then classically implement a kernel method (such as kernel ridge regression for example) using the chosen kernel. We refer to this approach as \textit{kernel-based dequantization}.

The first immediate obstacle to this approach, is that it can be unclear how to select an appropriate kernel from the set of candidates in a principled way. The second obstacle is that the candidate kernels are specified by exponentially large feature maps, which prevents the efficient evaluation of these kernels via naive classical algorithms -- i.e. it is not always clear how to evaluate these kernels without first constructing the feature vectors explicitly. In Ref.~\cite{shin2024} this latter obstacle is overcome by noticing that at least one candidate classical feature map is in fact a tensor product map, and therefore the associated classical kernel can be evaluated efficiently classically via tensor network techniques.
On the other hand, Refs.~\cite{landman2022classicallyapproximatingvariationalquantum,Sweke2023potential} have explored the use of Random Fourier Features (RFF) for \textit{approximating} the classical feature map with a randomly drawn feature map of smaller dimension. In particular, Ref.~\cite{Sweke2023potential} has identified necessary and sufficient conditions for efficient dequantization via RFF, in terms of both the candidate kernel and the problem to be solved, which helps to facilitate kernel selection. We stress that the conditions from~\cite{Sweke2023potential} also imply the existence of settings in which kernel based dequantization cannot be efficient. One can indeed construct an explicit example of such a setting by considering the problem of learning parities, which can be solved efficiently using a PQC-based model, but cannot be solved efficiently by any kernel method~\cite{Jerbi_2023}.

With this in mind, in this note we enhance the toolbox for kernel-based dequantization, by showing that a particular subset of candidate kernels for dequantization can in fact be evaluated exactly and efficiently classically. In this sense we provide a ``kernel trick'' for these kernels -- i.e. an efficient classical method for their evaluation which does not require explicitly constructing and taking the inner product of feature vectors. This method is directly inspired by the work of Ref.~\cite{shin2024}, and works by providing a matrix product state (MPS) representation for the feature map associated with the kernel, which then allows for the kernel to be evaluated through efficient contraction of a tensor network. This kernel trick allows for the efficient classical implementation of kernel methods with these kernels, without requiring any randomized approximations. As such, these kernels provide a natural set of candidate kernels for kernel-based dequantization, which can be used within a diverse range of kernel methods. 

We note that, interestingly, the set of kernels for which we provide a kernel trick is the same set of kernels which were shown in Ref.~\cite{Sweke2023potential} to be efficient to approximate within an RFF-based approach. As such, the insights from Ref.~\cite{Sweke2023potential} into the inductive bias and generalization capacity of such kernels, can still be used to inform kernel selection outside of the RFF-based approach. Additionally, while our results may seem to immediately remove the motivation for randomized kernel-based methods such as RFF, this is not necessarily the case, and we provide a discussion on the relative merits and disadvantages of RFF in light of the possibility of exact and efficient kernel evaluation.

\subsection{Structure of this work}\label{ss:structure}

We start in Section~\ref{s:prelim} by providing all the preliminaries necessary for this work. In particular, we start in Sections~\ref{s:PQC_QML} and~\ref{ss:PQC_linear}, with a brief overview of PQC models for variational QML, following closely the presentation of Ref.~\cite{Sweke2023potential}. With this in hand we then present in Section~\ref{ss:pqc_inspired} a set of kernels -- which we call ``PQC-inspired kernels'' -- whose reproducing kernel Hilbert spaces are guaranteed to contain all functions realizable by a specific PQC model. This then allows us to describe, in Section~\ref{ss:kernel_based_dequantization}, the kernel-based approach to dequantization of variational QML. With this in hand we review prior work on this approach in Section~\ref{ss:prior_work}, which allows us to state our contributions clearly in Section~\ref{ss:contributions}.  Finally, we present in Section~\ref{ss:data_to_freq} a variety of technical details concerning PQC models that will be needed for our main results. With this in hand, we then define in Section~\ref{s:MPS_reweight} the notion of a PQC-inspired kernel reweighted via weighting induced from a symmetric MPS. This allows us to show our main result in Section~\ref{s:derandomizing}, namely that  any PQC-inspired kernel reweighted via a weighting induced from a symmetric MPS can be evaluated exactly and efficiently classically. We then discuss in Section~\ref{ss:derandomize} the implications for the RFF-based approach to dequantization, in Section~\ref{s:ETK} the relation to the notion of ``Entangled Tensor Kernels"~\cite{shin2025newperspectivesquantumkernels}, and we conclude in Section~\ref{s:conclusions}.

\section{Preliminaries}\label{s:prelim}

\subsection{PQC-based models for variational QML}\label{s:PQC_QML}

We consider models for regression defined by a parameterized quantum circuit (PQC) $C$ and an observable $O$. The gates of $C$ can depend on data points $x\in\mathcal{X}\subseteq\mathbb{R}^d$ and components of a vector $\theta\in\Theta$ of variational parameters.  For each data point $x$ and each vector of variational parameters $\theta$, we denote the unitary realized by the circuit as $U(x,\theta)$. Additionally, to every $\theta\in \Theta$, we define the function $f_\theta:\mathbb{R}^d\rightarrow\mathbb{R}$ via
\begin{equation}
f_{\theta}(x) = \langle 0|U^{\dagger}(x,\theta)OU(x,\theta)|0\rangle,
\end{equation}
for all $x\in\mathcal{X}$. We refer to such functions as \textit{PQC models}. It is by now well known that, under mild assumptions on the structure of gates in the circuit $C$, these PQC models admit a succinct description given by
\begin{align}\label{eq:fourier_rep}
f_\theta(x) = \sum_{\omega\in\tilde{\Omega}} c_{\omega}(\theta) e^{i\langle\omega, x\rangle}.
\end{align}
In particular, one has that:
\begin{enumerate}
\item The set of frequency vectors $\tilde{\Omega}\subseteq\mathbb{R}^d$ is completely determined by the gates in $C$ which are parameterized by data-points $x\in\mathbb{R}^d$~\cite{Schuld_2021,gil2020input,Caro_2021}.
\item The frequency coefficients $c_{\omega}(\theta)$ depend on all the gates in the circuit, but not on $x$, and are constrained in some way which usually does not admit a concise expression~\cite{mhiri2024constrainedvanishingexpressivityquantum}. 
\end{enumerate}
Additionally, we know that $\omega_0 := (0,\ldots,0)\in\tilde{\Omega}$, and that the non-zero frequencies in $\tilde{\Omega}$ come in mirror pairs -- i.e. $\omega\in\tilde{\Omega}$ implies $-\omega\in\tilde{\Omega}$. Moreover, one can show that $c_\omega(\theta) = c_{-\omega}^*(\theta)$, which ensures that $f_\theta(x)$ is real. Finally, we note that given the structure of $\tilde{\Omega}$, we can perform an arbitrary splitting of pairs to  redefine $\tilde{\Omega} \coloneqq \Omega \cup \left(-\Omega\right)$, where $\Omega \cap \left(-\Omega\right) = \{\omega_0\}$. We define $M = | \Omega\setminus \{\omega_0\} |$ which allows us to write $\Omega = \{\omega_0,\omega_1,\ldots,\omega_{M}\}$ and $\tilde{\Omega} = \{-\omega_M,\ldots,-\omega_1,\omega_0,\omega_1,\ldots,\omega_M\}$. We will further discuss the structure of the set $\tilde{\Omega}$ in Section~\ref{ss:data_to_freq}.

\subsection{Linearity of PQC models}\label{ss:PQC_linear}

It is known that all PQC models are \textit{linear models} with respect to a feature map defined by $\Omega$. To see this, we first define
\begin{align}
a_\omega(\theta) &= c_\omega(\theta) + c_{-\omega}(\theta),\\
b_\omega(\theta) &= i(c_\omega(\theta) - c_{-\omega}(\theta)).
\end{align}
Using this, together with the structure of $\tilde{\Omega}$ discussed above, we can rewrite Eq.~\eqref{eq:fourier_rep} as
\begin{align}
f_\theta(x) &= c_{\omega_0}(\theta) + \sum_{i=1}^{M}\left(a_{\omega_i}(\theta)\cos(\langle\omega_i,x\rangle) + b_{\omega_i}(\theta)\sin(\langle\omega_i,x\rangle)\right)\\
&=\langle c(\theta),\phi_{\Omega}(x)\rangle,\label{eq:linear_map}
\end{align}
where we have defined 
\begin{align}
c(\theta) &= \left(c_{\omega_0}(\theta),a_{\omega_1}(\theta),b_{\omega_1}(\theta),\ldots,a_{\omega_M}(\theta),b_{\omega_M}(\theta) \right),\\
\phi_\Omega(x) &= \big(1,\cos(\langle\omega_1,x\rangle),\sin(\langle\omega_1,x\rangle),\ldots,\cos(\langle\omega_M,x\rangle),\sin(\langle\omega_M,x\rangle) \big).
\end{align}
As such, we see from Eq.~\eqref{eq:linear_map} that $f_\theta$ is indeed a linear function with respect to the feature map $\phi_\Omega:\mathbb{R}^d\rightarrow\mathbb{R}^{(2M + 1)}$.

\subsection{PQC-inspired kernels}\label{ss:pqc_inspired}

From any feature map $\phi:\mathbb{R}^d \rightarrow \mathbb{R}^{d'}$ and positive semidefinite matrix $A\in\mathbb{R}^{d'\times d'}$, one can define a valid kernel ${K_{(\phi,A)}:\mathbb{R}^d\times\mathbb{R}^d\rightarrow \mathbb{R}}$ via 
\begin{equation}
K_{(\phi,A)}(x,x') = \langle \phi(x),A\phi(x')\rangle.
\end{equation}
This follows from the fact that every positive semidefinite $A$ defines a valid inner product on the feature space $\mathbb{R}^{d'}$. Given some parameterized quantum circuit $C$, which leads to a frequency set $\Omega$,  we call the kernel $K_{(\phi_\Omega,A)}$ a \textit{PQC-inspired} kernel, for any positive semidefinite $A$. For simplicity, we denote this kernel with $K_{(\Omega,A)}$. We now define a special case, which will be of particular interest to us. To do this, let $w = (w_0,\ldots,w_M)\in\mathbb{R}^{(|\Omega|)}$ be any vector with all components non-zero, and define
\begin{equation}
A_w = \mathrm{diag}\left(\frac{1}{\|w\|_2}(w_0,w_1,w_1,\ldots,w_M,w_M)\right), 
\end{equation}
and $\tilde{A}_w = A_w^2$, which is positive definite by construction. We now note that
\begin{align}
K_{(\Omega,\tilde{A}_w)}(x,x') = \frac{1}{\|w\|_2^2}\sum_{i=0}^M w_i^2 \left[\cos(\langle \omega_i,x-x'\rangle)\right],
\end{align}
and therefore $K_{(\Omega,\tilde{A}_w)}$ is both \textit{normalized} (i.e. $K_{(\Omega,\tilde{A}_w)}(x,x) = 1$) and \textit{shift-invariant} (i.e. depends only on $x-x'$). These are convenient properties, which motivate this special case. We refer to $w$ as a reweighting of $\Omega$, and to the kernel $K_{(\Omega,\tilde{A}_w)}$ as the PQC-inspired kernel reweighted by $w$. When $w=(1,\ldots,1)$ we have $\tilde{A}_w\propto\mathbb{I}_{d'}$ and we call $K_{(\Omega,\tilde{A}_w)}$ the canonical PQC-inspired kernel. For simplicity we will denote $K_{(\Omega,\tilde{A}_w)}$ with $K_{(\Omega,w)}$. 

Finally, we note that if we define $\phi^{(1)}_{(\Omega,w)}:\mathbb{R}^d\rightarrow \mathbb{R}^{(2M+1)}$ and $\phi^{(2)}_{(\Omega,w)}:\mathbb{R}^d\rightarrow \mathbb{C}^{(2M+1)}$, via:
\begin{align}
|\phi^{(1)}_{(\Omega,w)}(x)\rangle &\coloneqq \frac{1}{\|w\|_2}\big( w_0,w_1\cos(\langle \omega_1, x\rangle), w_1\sin(\langle \omega_1, x\rangle),\ldots,w_{M}\cos(\langle \omega_{M}, x\rangle),w_{M}\sin(\langle \omega_{M}, x\rangle)\big),\\
|\phi^{(2)}_{(\Omega,w)}(x)\rangle &\coloneqq \frac{1}{\sqrt{2}\|w\|_2}\big(w_M e^{-i\langle\omega_M,x\rangle},\ldots,w_1 e^{-i\langle\omega_1,x\rangle}, \sqrt{2} w_0,w_1e^{i\langle\omega_1,x\rangle},\ldots, w_M e^{i\langle\omega_M,x\rangle}   \big),
\end{align}
Then one has
\begin{equation}
K_{(\Omega,w)}(x,x') = \langle \phi^{(1)}_{(\Omega,w)}(x),\phi^{(1)}_{(\Omega,w)}(x')\rangle = \langle \phi^{(2)}_{(\Omega,w)}(x),\phi^{(2)}_{(\Omega,w)}(x')\rangle.
\end{equation}
These alternative expressions will be of use to us in Section~\ref{s:derandomizing}.

\subsection{Kernel-based dequantization of variational QML}\label{ss:kernel_based_dequantization}

Assume some parameterized quantum circuit $C$, which leads to frequency set $\Omega$. For any \textit{positive definite} matrix $A$, if one uses a kernel-based learning algorithm (such as kernel ridge regression for example) with the PQC-inspired kernel $K_{(\phi_\Omega,A)}$, the output function will be a linear function with respect to the feature map $\phi_\Omega$. In fact, in many cases one can show that the output function will actually be the \textit{empirical risk minimizer} over all such linear functions~\cite{scholkopf2002learning,SVMbook}. As PQC-based optimization via $C$ will also output a linear function with respect to $\phi_\Omega$ (typically without any additional guarantees) a natural approach to dequantizing variational QML based on PQC optimization via $C$, is the following:

\begin{center}
\textit{Replace PQC optimization via $C$ with a kernel-based method using a kernel $K_{(\Omega,A)}$.}
\end{center}

We refer to such an approach as \textit{kernel-based dequantization}. Unfortunately, there are two immediate obstacles to this approach:

\begin{enumerate}
\item \textbf{Efficiency:} As we will see in Section~\ref{ss:data_to_freq}, the size of the frequency set $\Omega$ often scales \textit{exponentially} with respect to $d$. As such, it is not a-priori clear whether one can evaluate $K_{(\Omega,A)}$ efficiently classically, which is a necessary condition for the efficient implementation of many kernel-based learning algorithms.
\item \textbf{Generalization:} Ultimately the performance of a hypothesis should be measured by its performance on unseen data -- i.e. by its generalization performance -- and a guarantee on the empirical risk of an output hypothesis is not on its own sufficient to prove a guarantee on generalization. In particular, it is not on its own sufficient to guarantee that the output hypothesis generalizes better than the hypothesis output from PQC optimization.
\end{enumerate}

In light of the above obstacles, given a paramterized quantum circuit $C$ leading to frequency set $\Omega$, the first step towards practical implementation of kernel based dequantization of $C$ is the identification of a kernel method, and kernel $K_{(\Omega,A)}$ which can be implemented efficiently. Additionally, in order for such a method to \textit{provably} dequantize PQC based optimization, one then also needs to prove a relative generalization bound, which proves that the output of the kernel-based method generalizes at least as well as the output of PQC-based optimization.

\subsection{Prior work on kernel-based dequantization}\label{ss:prior_work}

The kernel-based approach to dequantization was first proposed and explored in Ref.~\cite{landman2022classicallyapproximatingvariationalquantum}, who noticed that the kernel-based method of Random Fourier Features (RFF) -- in which kernel ridge regression via some kernel $K$ is replaced by linear regression with respect to a low-dimensional feature map with components randomly selected from the eigenspectrum of $K$ -- can sometimes be efficiently implemented even when $K$ cannot be evaluated efficiently. As such, this offers a promising direction for kernel-based dequantization. However, this original work of Ref.~\cite{landman2022classicallyapproximatingvariationalquantum} focused primarily on the canonical PQC-inspired kernel, and implicitly considered \textit{universal} PQC-models when providing analytical guarantees. Ref.~\cite{Sweke2023potential} then extended this line of work, showing that one can in fact consider any reweighted PQC-inspired kernel, and provided necessary and sufficient conditions for both efficiency and good generalization of RFF-based dequantization, in terms of the circuit $C$, the reweighting vector $w$, and the problem which is being addressed.

Another approach was taken in Ref.~\cite{shin2024}, who showed via a tensor-network analysis, that for every parameterized quantum circuit architecture $C$, one can identify a kernel which \textit{can} be evaluated exactly and efficiently classically, and is also guaranteed to output a linear function with respect to the feature map $\phi_\Omega$. As such, this kernel also provides a natural starting point for kernel-based dequantization, and one only needs to prove suitable generalization bounds.

\subsection{Our contribution}\label{ss:contributions}

In light of the above mentioned work, our primary contribution here is the following:

\begin{center}\textit{We show that for a wide class of reweightings $w$, the reweighted PQC-inspired kernels $K_{(\Omega,w)}$ can in fact be evaluated exactly and efficiently classically.}
\end{center}

More specifically, we show the above result for reweightings that are induced from symmetric matrix product states -- a class of reweightings that we define rigorously in Section~\ref{s:MPS_reweight}. As such, these kernels provide natural candidates for kernel-based dequantization, as one can immediately use a wide variety of kernel based methods, without requiring any randomized approximations. In another sense, we provide a \textit{kernel-trick} for reweighted PQC-inspired kernels $K_{(\Omega,w)}$ with reweightings induced from symmetric MPS -- i.e. a way to evaluate these kernels exactly and efficiently classically without having to explicitly construct and take the inner product of feature vectors.

\subsection{From data-encoding strategy to frequency sets}\label{ss:data_to_freq}

In this work, we consider an important sub-class of parametrized quantum circuits in which the classical data $x$ enters only via Hamiltonian time evolutions, whose duration is controlled by a single component of $x$. To be more precise, for $x = (x_1,\ldots,x_d)\in\mathbb{R}^d$, we assume that each gate in the circuit $C$ which depends on $x$ is of the form
\begin{equation}
V_{(j,k)}(x_j) = e^{-iH^{(j)}_kx_j},
\end{equation}
for some Hamiltonian $H^{(j)}_k$. We stress that we exclude here more general encoding schemes, such as those which allow for time evolutions of parameterized linear combinations of Hamiltonians. We denote by $\mathcal{D}^{(j)} = \{H^{(j)}_k\,|\, k\in [L_j]\}$ the set of all $L_j$ Hamiltonians which are used to encode the component $x_j$ at some point in the circuit, and we call the tuple
\begin{equation}
\mathcal{D} \coloneqq \left(\mathcal{D}^{(1)},\ldots,\mathcal{D}^{(d)}\right)
\end{equation}
the \textit{data-encoding strategy}. For any such parameterized quantum circuit $\mathcal{C}$ the frequency set $\tilde{\Omega}$ appearing in Eq.~\eqref{eq:fourier_rep} depends only on the data-encoding strategy $\mathcal{D}$~\cite{Schuld_2021}. To denote that we are in this specific setting, we denote this frequency set as $\tilde{\Omega}_\mathcal{D}$. As discussed at the end of Section~\ref{s:PQC_QML}, we can introduce an arbitrary splitting of mirror pairs in $\tilde{\Omega}_\mathcal{D}$, so that $\tilde{\Omega}_\mathcal{D} \coloneqq\Omega_{\mathcal{D}}\cup \left(-\Omega_{\mathcal{D}}\right)$, where $\Omega_{\mathcal{D}}\cap\left(-\Omega_{\mathcal{D}}\right) = \{\omega_0\}$

We refer to Refs.~\cite{Schuld_2021,Sweke2023potential} for a fully detailed exposition on the construction and structure of the frequency set $\tilde{\Omega}_\mathcal{D}$. However, we mention one important fact necessary to us here. Specifically, we know that
\begin{equation}\label{eq:cartesian_product}
\tilde{\Omega}_\mathcal{D} = \tilde{\Omega}^{(1)}_\mathcal{D}\times \ldots\times\tilde{\Omega}^{(d)}_\mathcal{D},
\end{equation}
where $\tilde{\Omega}^{(j)}_\mathcal{D}\subset \mathbb{R}$ depends only on $\mathcal{D}^{(j)}$, the data-encoding strategy for data component $j$. In particular,  $\tilde{\Omega}_\mathcal{D}$ has a \textit{Cartesian product structure}. If we define $\tilde{M}_j := |\tilde{\Omega}^{(j)}_\mathcal{D}|$, and make the assumption that $\tilde{M}_j$ is independent of $d$, then we immediately see that 
\begin{equation}
|\tilde{\Omega}_\mathcal{D}| \geq \tilde{M}_{\text{min}}^d
\end{equation}
where $\tilde{M}_{\text{min}} = \min\{\tilde{M}_j\}$ -- i.e.,  we see that the number of frequencies in $\tilde{\Omega}_\mathcal{D}$ scales \textit{exponentially} with respect to $d$. It follows from the discussion in Section~\ref{s:PQC_QML} on the structure of $\tilde{\Omega}_\mathcal{D}$, that for all $j$ one must have $0\in \tilde{\Omega}^{(j)}_\mathcal{D}$, and that all the non-zero frequencies in $\tilde{\Omega}^{(j)}_\mathcal{D}$ come in mirror pairs -- i.e. that $\omega\in\tilde{\Omega}^{(j)}_\mathcal{D}$ implies $-\omega\in\tilde{\Omega}^{(j)}_\mathcal{D}$. Using this, and defining $\tilde{M_j} = 2M_j +1$, we can choose to label the elements of $\tilde{\Omega}^{(j)}_\mathcal{D}$ as
\begin{align}
\tilde{\Omega}^{(j)}_\mathcal{D} &= \{\omega^{(j)}_{-M_j},\ldots,\omega^{(j)}_{-1},\omega^{(j)}_{0},\omega^{(j)}_{1},\ldots,\omega^{(j)}_{M_j}\}\\
&=\{-\omega^{(j)}_{M_j},\ldots,-\omega^{(j)}_{1},0,\omega^{(j)}_{1},\ldots,\omega^{(j)}_{M_j}\}.
\end{align}
As a consequence of the cartesian product structure of $\tilde{\Omega}_\mathcal{D}$ we then have
\begin{equation}\label{eq:freq_set_labelling}
\tilde{\Omega}_\mathcal{D} = \left\{\omega_{k_1,\ldots,k_d} = \left((\omega^{(1)}_{k_d},\ldots,\omega^{(d)}_{k_1}\right)\,\big\vert\, k_1\in \{-M_1,\ldots,M_1\}\,,\ldots, k_d\in\{-M_d,\ldots,M_d\}\right\}.
\end{equation}

\section{Reweightings induced from symmetric Matrix Product States (MPS)}\label{s:MPS_reweight}

Consider a frequency set $\Omega_\mathcal{D}$, derived from a data-encoding strategy $\mathcal{D}$, as discussed in Section~\ref{ss:data_to_freq}. Recall from Section~\ref{ss:pqc_inspired}, that a reweighting of $\Omega_\mathcal{D}$ is some $w\in\mathbb{R}^{|\Omega_\mathcal{D}|}$ with all components strictly non-zero. In this section we define the notion of a reweighting $w\in\mathbb{R}^{|\Omega_\mathcal{D}|}$ \textit{induced from a symmetric matrix product state (MPS)}. In Section~\ref{s:derandomizing} we will then show our main result, namely that the reweighted PQC-inspired kernel $K_{(\Omega_\mathcal{D},w)}$ can be evaluated exactly and efficiently classically, whenever $w$ is a reweighting of $\Omega_\mathcal{D}$ induced from a symmetric MPS. In this section we assume familiarity with tensor network notation, for an introduction we refer to Ref.~\cite{Bridgeman_2017}. Additionally, for simplicity, from this point on we will simply write ``weighting'' instead of ``reweighting''.

To start, assume a data-encoding strategy $\mathcal{D}$, as per Section~\ref{ss:data_to_freq}, which implies a frequency set $\tilde{\Omega}_\mathcal{D}$. Recall that $\omega\in\tilde{\Omega}_\mathcal{D}$ implies that $-\omega\in\tilde{\Omega}_\mathcal{D}$ and that $\Omega_\mathcal{D}$ is then defined by an arbitrary splitting of mirror pairs from $\tilde{\Omega}_\mathcal{D}$. We define a weighting of $\tilde{\Omega}_\mathcal{D}$ as any $\tilde{w}\in\mathbb{R}^{|\tilde{\Omega}_\mathcal{D}|}$ with all components strictly non-zero. It is helpful for us to think of such a weighting as a function $\tilde{w}:\tilde{\Omega}_\mathcal{D}\rightarrow \mathbb{R}$. Analogously, we will think of a weighting of $\Omega_\mathcal{D}$ as a function $w:\Omega_\mathcal{D}\rightarrow \mathbb{R}$. Given a weighting $\tilde{w}$ of $\tilde{\Omega}_\mathcal{D}$, we define the weighting $w$ of $\Omega_\mathcal{D}$ \textit{induced} from $\tilde{w}$ via
\begin{equation}
w[\omega] = \tilde{w}[\omega]
\end{equation}
for all $\omega\in\Omega_\mathcal{D}$. Note that it may be the case that $\tilde{w}[\omega]\neq\tilde{w}[-\omega]$. With this in mind, we call any weighting $\tilde{w}$ of $\tilde{\Omega}_\mathcal{D}$ which also satisfies $\tilde{w}[\omega]=\tilde{w}[-\omega]$ for all $\omega\in\Omega_\mathcal{D}$ a \textit{symmetric} weighting of $\tilde{\Omega}_\mathcal{D}$. 

We now want to define the notion of an MPS weighting of $\tilde{\Omega}_\mathcal{D}$. To this end, note that as a consequence of Eq.~\eqref{eq:freq_set_labelling}, any weighting $\tilde{w}$ of $\tilde{\Omega}_\mathcal{D}$ can be represented as a $d$-index tensor where the $j$'th index is $2M_j+1$ dimensional (running from $-M_j$ to $M_j$). More specifically, we have
\begin{equation}
\tilde{w}[\omega_{k_1,\ldots,k_d}] = \vcenter{\hbox{\includegraphics{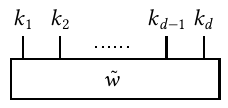}}}.
\end{equation}
We know that any such tensor can be decomposed into an MPS -- i.e. we can always write
\begin{equation}
\tilde{w}[\omega_{k_1,\ldots,k_d}] = \vcenter{\hbox{\includegraphics{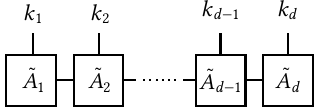}}}.
\end{equation}
We will call a weighting $\tilde{w}$ of $\tilde{\Omega}_\mathcal{D}$ an MPS weighting if the MPS decomposition of $\tilde{w}$ has bond-dimension polynomial in $d$. Given all of this, we simply say that a weighting $w$ of $\Omega_\mathcal{D}$ is \textit{induced from a symmetric MPS} if it is induced from a weighting $\tilde{w}$ of $\tilde{\Omega}_\mathcal{D}$ which is both an MPS weighting and symmetric.

Finally, we note that given any MPS weighting $\tilde{w}$, we can straightforwardly obtain a symmetric MPS weighting $\tilde{w}_s$. To do this, we start by defining for each $j\in[d]$ the "index-flip operator" $\pi:\mathbb{Z}\rightarrow\mathbb{Z}$ which acts via $\pi(k) = -k$. We then have
\begin{align}
\tilde{w}_s[\omega_{k_1,\ldots,k_d}] 
&=\vcenter{\hbox{\includegraphics{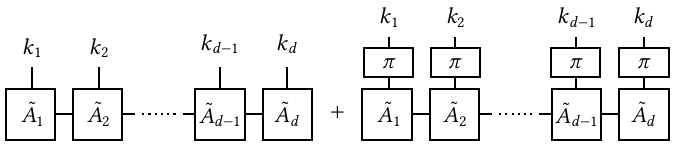}}},\\
&:= \vcenter{\hbox{\includegraphics{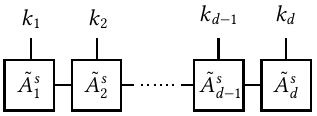}}},
\end{align}
which can be easily checked to be appropriately symmetrized and of at most double the bond-dimension as $\tilde{w}$. From this point on, we will use the notation $\tilde{w}_s$ to indicate a weighting of $\tilde{\Omega}_\mathcal{D}$ which is symmetric, and the notation $\tilde{A}^s_i$ to denote the $i$'th MPS-tensor of a symmetric MPS weighting (whether or not it arises from a symmetrization procedure).

\section{Exact and efficient evaluation of kernels with MPS-induced reweightings}\label{s:derandomizing}

In the previous section we have defined the notion of a weighting of $\Omega_\mathcal{D}$ induced from a symmetric MPS. In this section we will show that for weightings $w$ of $\Omega_\mathcal{D}$ induced from a symmetric MPS, the associated PQC-inspired kernel $K_{(\Omega_\mathcal{D},w)}$ can in fact be evaluated exactly and efficiently classically! Given this, such kernels provide natural candidates for kernel-based dequantization algorithms. For ease of notation, in this section we write $K_{(\mathcal{D},w)}$ instead of $K_{(\Omega_\mathcal{D},w)}$ and $\phi^{(i)}_{(\mathcal{D},w)}$ instead of $\phi^{(i)}_{(\Omega_\mathcal{D},w)}$.

Recall from Section~\ref{ss:pqc_inspired} that one has
\begin{equation}
K_{(\mathcal{D},w)}(x,x') = \langle \phi^{(2)}_{(\mathcal{D},w)}(x)|\phi^{(2)}_{(\mathcal{D},w)}(x')\rangle,
\end{equation}
where in the multi-index notation of the Sections~\ref{ss:data_to_freq} and~\ref{s:MPS_reweight} we can write $|\phi^{(2)}_{(\mathcal{D},w)}(x')\rangle$ as
\begin{align}
|\phi^{(2)}_{(\mathcal{D},w)}(x)\rangle
&= \frac{1}{\sqrt{2}\|w\|_2}
\begin{pmatrix}
    w(\omega_{M_1,\ldots,M_d})e^{-i\langle\omega_{M_1,\ldots,M_d},x\rangle}\\
    \vdots \\
    w(\omega_{k_1,\ldots,k_d}) e^{-i\langle\omega_{k_1,\ldots,k_d},x\rangle}\\
    \vdots\\
    \sqrt{2} w(\omega_{0,\ldots,0})\\
    \vdots\\
    w(\omega_{k_1,\ldots,k_d}) e^{i\langle\omega_{k_1,\ldots,k_d},x\rangle}\\
    \vdots \\
    w(\omega_{M_1,\ldots,M_d})e^{i\langle\omega_{M_1,\ldots,M_d},x\rangle}
\end{pmatrix}.
\end{align}
Our strategy is to show that for weightings $w$ induced from symmetric MPS, the feature map $|\phi^{(2)}_{(\mathcal{D},w)}(x)\rangle$ also has a polynomial bond-dimension MPS representation. To this end, we start by defining
\begin{align}
|\psi^{(j)}_{\mathcal{D}}(x)\rangle = \vcenter{\hbox{\includegraphics{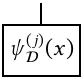}}} = \begin{pmatrix} e^{i\omega^{(1)}_{-M_j} x_j}\\\vdots\\e^{i\omega^{(1)}_{-1} x_j} \\e^{i\omega^{(1)}_{0} x_j}\\e^{i\omega^{(1)}_{1} x_j} \\\vdots\\e^{i\omega^{(1)}_{M_j} x_j}\end{pmatrix}= \begin{pmatrix} e^{-i\omega^{(1)}_{M_j} x_j}\\\vdots\\e^{-i\omega^{(1)}_{1} x_j} \\1\\e^{i\omega^{(1)}_{1} x_j} \\\vdots\\e^{i\omega^{(1)}_{M_j} x_j}\end{pmatrix},
\end{align}
for all $x\in\mathbb{R}^d$ and for all $j\in\{1,\ldots,d\}$. We then have 
\begin{align}
|\psi_\mathcal{D}^{(1)}(x)\rangle\otimes\ldots\otimes |\psi_\mathcal{D}^{(d)}(x)\rangle =\vcenter{\hbox{\includegraphics{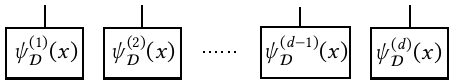}}} = 
 \begin{pmatrix}
    e^{-i\langle\omega_{M_1,\ldots,M_d},x\rangle}\\
    \vdots \\
     e^{-i\langle\omega_{k_1,\ldots,k_d},x\rangle}\\
    \vdots\\
    1\\
    \vdots\\
    e^{i\langle\omega_{k_1,\ldots,k_d},x\rangle}\\
    \vdots \\
    e^{i\langle\omega_{M_1,\ldots,M_d},x\rangle}
\end{pmatrix} 
\end{align}
This is almost $|\phi^{(2)}_{(\mathcal{D},w)}(x)\rangle$, we just need to introduce the weighting of each vector component. To this end, we first define the $d$-index tensor $C$ via
\begin{equation}
C_{k_1,\ldots,k_d}= \begin{cases} 1&\text{ if } (k_1,\ldots,k_d)\neq (0,\ldots, 0)\\
\sqrt{2}&\text{ if } (k_1,\ldots,k_d)= (0,\ldots, 0)
\end{cases}.
\end{equation}
The tensor $C$ admits a representation as a matrix product state with bond-dimension 2, so that we can write
\begin{equation}
C_{k_1,\ldots,k_d} = \vcenter{\hbox{\includegraphics{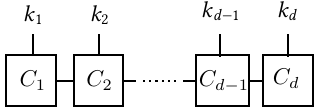}}}.
\end{equation}
Next, we define the delta, or ``copy'' tensor
\begin{equation}
\vcenter{\hbox{\includegraphics{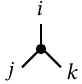}}} = \delta_{i,j,k}.
\end{equation}
Now, assume that $w$ is a weighting of $\Omega_\mathcal{D}$ induced from a symmetric MPS weighting $\tilde{w}_s$ of $\tilde{\Omega}_\mathcal{D}$. This then allows us to define the $d$-index tensor $B^{\tilde{w}_s}$ via
\begin{align}
B^{\tilde{w}_s}_{k_1,\ldots,k_d} &= C_{k_1,\ldots,k_d} \times \tilde{w}_s[\omega_{k_1,\ldots,k_d}]\\
&= \vcenter{\hbox{\includegraphics{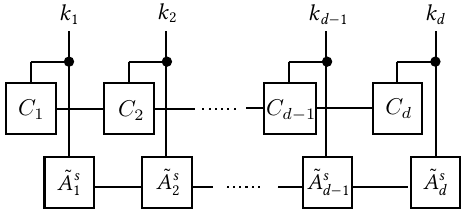}}}\\
&= \vcenter{\hbox{\includegraphics{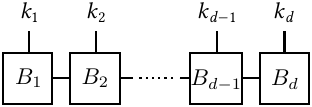}}},
\end{align}
which by construction satisfies
\begin{equation}
B^{\tilde{w}_s}_{k_1,\ldots,k_d} = \begin{cases} 
\tilde{w}_s[\omega_{k_1,\ldots,k_d}]&\text{ if } (k_1,\ldots,k_d)\neq (0,\ldots,0)\\
\sqrt{2}\tilde{w}_s[\omega_{k_1,\ldots,k_d}]&\text{ if } (k_1,\ldots,k_d)= (0,\ldots,0)
\end{cases}
\end{equation}
We stress that because $\tilde{w}^s$ is a symmetric weighting of $\tilde{\Omega}_\mathcal{D}$ we also have $B^{\tilde{w}_s}_{k_1,\ldots,k_d} = B^{\tilde{w}_s}_{-k_1,\ldots,-k_d}$. Using this, and the $\sqrt{2}$ factor for the weight associated with $(k_1,\ldots,k_d)= (0,\ldots,0)$, one can verify that
\begin{align}
\vcenter{\hbox{\includegraphics{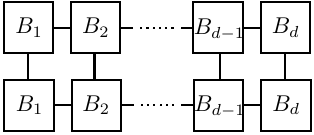}}} &= \sum_{k_1 = -M_1}^{M_1}\ldots\sum_{k_d = -M_d}^{M_d} \left(B^{\tilde{w}_s}_{k_1,\ldots,k_d}\right)^2,\\
&= \left(\sqrt{2}\tilde{w}_s[\omega_{0,\ldots,0}]\right)^2 + 2\sum_{\omega\in\Omega_\mathcal{D}\setminus \omega_{0,\ldots,0}}(\tilde{w}_s[\omega])^2\\
&= 2(w_s[\omega_{0,\ldots,0}])^2 + 2\sum_{\omega\in\Omega_\mathcal{D}\setminus \omega_{0,\ldots,0}}(w_s[\omega])^2\\
&= 2\|w\|^2_2. 
\end{align}
One can then see that 
\begin{align}
|\phi^{(2)}_{(\mathcal{D},w)}(x)\rangle &= \frac{1}{\sqrt{2}\|w\|_2} \left( \vcenter{\hbox{\includegraphics{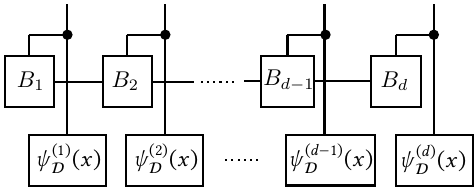}}}\right)\\
&:=\frac{1}{\sqrt{2}\|w\|_2} \left(\vcenter{\hbox{\includegraphics{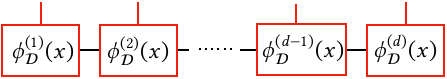}}}\right)\label{eq:MPS_rep}
\end{align}
which is the desired matrix product state representation that we were looking for. This then gives
\begin{align}
K_{(\mathcal{D},w)}(x,x') &= \frac{1}{2\|w\|^2_2} \left(\vcenter{\hbox{\includegraphics{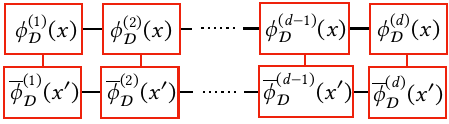}}}\right) \\
&=\left(\frac{\includegraphics{NewFigs/TNL14.pdf}}{\includegraphics{NewFigs/TNL15.pdf}}\right)\label{eq:K_final}
\end{align}

With the above established, let us now examine the complexity of evaluating $K_{(\mathcal{D},w)}(x,x')$ via contraction of Eq.~\eqref{eq:K_final}. To this end:
\begin{enumerate}
\item Recall that $|\psi^{(j)}_\mathcal{D}(x)\rangle$ is an $\tilde{M}_j = 2M_j+1$ dimensional vector. As before, we make the assumption that $M_{j}$ is a constant independent of $d$ for all $j$, and we now additionally define $\tilde{M}_{\max} = \max\{\tilde{M}_j\}$.
\item Assume that the symmetric MPS weighting $\tilde{w}_s$ of $\tilde{\Omega}_\mathcal{D}$ has bond-dimension $D$, which was already assumed to be polynomial in $d$.
\end{enumerate}
With the above, one can then see that the complexity of evaluating $K_{(\mathcal{D},w)}(x,x')$ via Eq.~\eqref{eq:K_final} is $O(dD^3\tilde{M}_{\max})$. More specifically, this follows from the standard ``edge-bond-bond'' contraction scheme for the taking the inner product of two MPS~\cite{Schollwock_2011,Bridgeman_2017}, together with consideration of the internal structure of each tensor, which adds constant factors to the complexity.

In light of the above, we can now make the following observation, which is the central result of this note:

\vspace{-5pt}
\begin{observation}[Efficient evaluation of PQC-inspired kernels reweighted via weightings induced from symmetric MPS]\label{obs:eff_contract} Given some data-encoding strategy $\mathcal{D}$, let $w$ be a weighting of $\Omega_{\mathcal{D}}$ induced from a symmetric MPS weighting $\tilde{w}^s$ of $\tilde{\Omega}_\mathcal{D}$. Then the reweighted PQC-inspired kernel $K_{(\mathcal{D},w)}$ can be evaluated exactly and efficiently classically.
\end{observation}
\vspace{-5pt}
As mentioned in Section~\ref{ss:contributions} note that one can interpret Eq.~\eqref{eq:K_final} as describing a \textit{kernel trick} for the evaluation of reweighted classically inspired-PQC kernels -- i.e. a method to evaluate the kernel without firstly fully constructing the feature vectors and then taking the inner product. As per Observation~\ref{obs:eff_contract}, this kernel trick is efficient when the reweighting is induced from a symmetric MPS (of polynomial bond-dimension)

\section{Implications for RFF-based dequantization}\label{ss:derandomize}

Given the results of Section~\ref{s:derandomizing}, one can straightforwardly and efficiently implement a variety of kernel-based methods for any PQC-inspired kernel $K_{(\mathcal{D},w)}$ reweighted via a weighting induced from a symmetric MPS. As such, one can think of these results as providing a means for simplifying or ``derandomizing'' RFF-based dequantization, by allowing one to simply implement a chosen kernel method, without requiring randomized approximations. A natural question is then following:

\begin{center}
    \textit{Are there any advantages to RFF-based dequantization?}
\end{center}

While we do not provide any definitive answers here, we do provide below a variety of perspectives on this question, which hopefully help to analyze the relative merits of RFF-based dequantization for concrete settings.

\textbf{Efficiency with respect to data-set size.} The original motivation for the introduction of Random Fourier Features was an improved complexity with respect to data-set size~\cite{rahimi2007random}. More specifically, let us consider a data-set of size~$n$ and assume some kernel $K$ which can be both evaluated in constant time, and for which one can sample from the distribution defined by its eigenspectrum in constant time. Then, implementing kernel ridge regression has space and time cost $O(n^2)$ and $O(n^3)$ respectively, where as RFF-based regression via $K$ has space and time cost $O(nS)$ and $(nS^2 + S^3)$, where $S$ are the number of samples used in the RFF-based algorithm to approximate $K$. As such, RFF based dequantization can be more efficient in the regime when $n > S$. Of course, in practice one also has to take into account the complexity of evaluating the kernel (which affects the complexity of kernel ridge regression), and of sampling from the kernel's eigenspectrum (which affects the complexity of RFF).

\textbf{Potentially efficient implementation where no kernel tricks exist.} In order for RFF to be efficient, one needs to be able to efficiently sample from a distribution defined by the eigenspectrum of the kernel. One might think that whenever this can be done efficiently, the kernel itself can be evaluated efficiently. Indeed, in Ref.~\cite{Sweke2023potential} it was shown that this sampling can be done efficiently for reweighted PQC-inspired kernels whenever the weighting is induced from a symmetric MPS, and here we have shown that precisely this set of kernels can also be efficiently evaluated classically! However, there does exist the possibility of PQC-inspired kernels $K_{(\mathcal{D},A)}$ for which the kernel cannot be efficiently evaluated, but for which one can efficiently sample from the eigenspectrum of the kernel, and therefore efficiently implement RFF. We leave identification of such kernels for future work.

\textbf{Generalization.} As discussed in Section~\ref{ss:kernel_based_dequantization},  good generalization bounds are required for any kernel-based approach to dequantization, if one wants to provide \textit{provable} guarantees of dequantization. To this end, Ref.~\cite{Sweke2023potential} has shown how can straightforwardly adapt known generalization bounds for RFF to the setting of reweighted PQC-inspired kernels, which then allows one to state rigorous necessary and sufficient conditions (on the reweighting $w$ and problem) for provable RFF-based dequantization. However, we note that one could similarly apply standard generalization bounds for algorithms such as kernel ridge regression~\cite{Canatar2021spectral} to reweighted PQC-inspired kernels, using insights into the spectrum of these kernels derived in Ref.~\cite{Sweke2023potential}. We leave a detailed comparison of generalization performance to future work.

\section{Relation to Entangled Tensor Kernels}\label{s:ETK}

Recently Ref.~\cite{shin2025newperspectivesquantumkernels} has introduced the notion of an \textit{entangled tensor kernel} (ETK) -- a generalization of a product kernel -- and shown that all embedding quantum kernels are ETKs. Here we note that ETKs also provide a suitable formalism for the description of reweighted PQC-inspired kernels. More specifically, we show that whenever $w$ is an MPS-induced weighting, then $K_{(\mathcal{D},w)}$ is in fact an ETK with a diagonal MPO core-tensor. To see this, we simply note that:
\begin{align}
\vcenter{\hbox{\includegraphics{NewFigs/TNL14.pdf}}} &= \vcenter{\hbox{\includegraphics{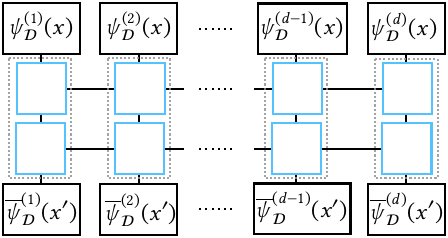}}}\\
&=\vcenter{\hbox{\includegraphics{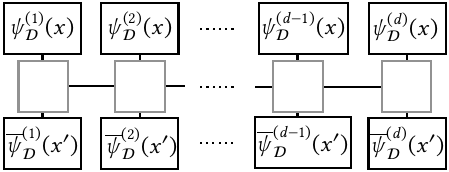}}}.
\end{align}
As such, by referring to the definition of an ETK in Ref.~\cite{shin2025newperspectivesquantumkernels} we see that $K_{(\mathcal{D},w)}$ is indeed an ETK, with local feature maps $\psi^{(j)}_\mathcal{D}$, and an MPO core tensor whose bond dimension is determined by the bond-dimension of the reweighting~$w$.

\section{Conclusions}\label{s:conclusions}

A natural approach to dequantization of PQC-based optimization is that of kernel-based dequantization, in which one replaces PQC optimization with a kernel method, using a PQC-inspired kernel. One obstacle to this approach, is that it is a-priori not clear whether PQC-inspired kernels can be evaluated efficiently. This has motivated the investigation of kernel-based dequantization based on the method of random Fourier features~\cite{landman2022classicallyapproximatingvariationalquantum,Sweke2023potential}, in which one makes use of randomized approximations of the kernel, which circumvent the need for evaluating the kernel exactly. In this work, we have shown that the class of reweighted PQC-inspired kernels, with reweightings induced from symmetric MPS, can in fact be evaluated exactly and efficiently classically. This ``kernel trick'' shows that for these kernels, the kernel-based approach to dequantization can be efficient, without any randomized approximations. Furthermore, as the spectral properties of reweighted PQC-inspired kernels are well understood~\cite{Sweke2023potential}, one can apply known generalization bounds, such as those for kernel ridge regression~\cite{Canatar2021spectral}, to obtain generalization bounds for the kernel-based approach to dequantization via reweighted PQC-inspired kernels. These results enrich the possible approaches to kernel-based dequantization, and allow for a comparison of existing methods for concrete situations and settings.

\section*{Acknowledgments}

RS is grateful for helpful and inspiring discussions concerning RFF-based dequantization with Mina Doosti and Mario Herrero-Gonz\'{a}lez. RS is also grateful to the Alexander von Humboldt Foundation for support, via the German Research Chair program. SS acknowledges support by the education and training program of the Quantum Information Research Support Center, funded through the National research foundation of Korea (NRF) by the Ministry of science and ICT (MSIT) of the Korean government (No.2021M3H3A1036573).
EGF is a 2023 Google PhD Fellowship recipient and acknowledges support by the Einstein Foundation (Einstein Research Unit on Quantum Devices), BMBF (Hybrid), and BMWK (EniQmA).

\printbibliography

@article{shin2024,
  title = {Dequantizing quantum machine learning models using tensor networks},
  author = {Shin, Seongwook and Teo, Yong Siah and Jeong, Hyunseok},
  journal = {Phys. Rev. Res.},
  volume = {6},
  issue = {2},
  pages = {023218},
  numpages = {24},
  year = {2024},
  month = {May},
  publisher = {American Physical Society},
  doi = {10.1103/PhysRevResearch.6.023218},
  url = {https://link.aps.org/doi/10.1103/PhysRevResearch.6.023218}
}

@book{SVMbook,
  title={Support vector machines},
  author={Steinwart, Ingo and Christmann, Andreas},
  year={2008},
  publisher={Springer Science \& Business Media},
  DOI={https://doi.org/10.1007/978-0-387-77242-4}
}

@misc{scholkopf2002learning,
  title={Learning with kernels: support vector machines, regularization, optimization, and beyond},
  author={Sch{\"o}lkopf, B},
  year={2002},
  publisher={The MIT Press}
}

@article{Canatar2021spectral, 
year = {2021}, 
title = {{Spectral bias and task-model alignment explain generalization in kernel regression and infinitely wide neural networks}}, 
author = {Canatar, Abdulkadir and Bordelon, Blake and Pehlevan, Cengiz}, 
journal = {Nature Communications}, 
doi = {10.1038/s41467-021-23103-1}, 
pmid = {34006842}, 
pmcid = {PMC8131612}, 
eprint = {2006.13198}, 
pages = {2914}, 
number = {1}, 
volume = {12}
}

@misc{Sweke2023potential,
      title={Potential and limitations of random Fourier features for dequantizing quantum machine learning}, 
      author={Ryan Sweke and Erik Recio and Sofiene Jerbi and Elies Gil-Fuster and Bryce Fuller and Jens Eisert and Johannes Jakob Meyer},
      year={2023},
      eprint={2309.11647},
      archivePrefix={arXiv},
      primaryClass={quant-ph}
}

@misc{landman2022classicallyapproximatingvariationalquantum,
      title={Classically Approximating Variational Quantum Machine Learning with Random Fourier Features}, 
      author={Jonas Landman and Slimane Thabet and Constantin Dalyac and Hela Mhiri and Elham Kashefi},
      year={2022},
      eprint={2210.13200},
      archivePrefix={arXiv},
      primaryClass={quant-ph},
      url={https://arxiv.org/abs/2210.13200}, 
}

@article{Schuld_2021,
   title={Effect of data encoding on the expressive power of variational quantum-machine-learning models},
   volume={103},
   ISSN={2469-9934},
   url={http://dx.doi.org/10.1103/PhysRevA.103.032430},
   DOI={10.1103/physreva.103.032430},
   number={3},
   journal={Physical Review A},
   publisher={American Physical Society (APS)},
   author={Schuld, Maria and Sweke, Ryan and Meyer, Johannes Jakob},
   year={2021},
   month=mar }

@article{gil2020input,
  title={Input redundancy for parameterized quantum circuits},
  author={Gil Vidal, Francisco Javier and Theis, Dirk Oliver},
  journal={Frontiers in Physics},
  volume={8},
  pages={297},
  year={2020},
  publisher={Frontiers Media SA}
}

@article{Jerbi_2023,
   title={Quantum machine learning beyond kernel methods},
   volume={14},
   ISSN={2041-1723},
   url={http://dx.doi.org/10.1038/s41467-023-36159-y},
   DOI={10.1038/s41467-023-36159-y},
   number={1},
   journal={Nature Communications},
   publisher={Springer Science and Business Media LLC},
   author={Jerbi, Sofiene and Fiderer, Lukas J. and Poulsen Nautrup, Hendrik and Kübler, Jonas M. and Briegel, Hans J. and Dunjko, Vedran},
   year={2023},
   month=jan }

@book{schuld2018supervised,
  title={Supervised learning with quantum computers},
  author={Schuld, Maria},
  year={2018},
  publisher={Springer}
}

@article{cerezo2021variational,
  title={Variational quantum algorithms},
  author={Cerezo, Marco and Arrasmith, Andrew and Babbush, Ryan and Benjamin, Simon C and Endo, Suguru and Fujii, Keisuke and McClean, Jarrod R and Mitarai, Kosuke and Yuan, Xiao and Cincio, Lukasz and others},
  journal={Nature Reviews Physics},
  volume={3},
  number={9},
  pages={625--644},
  year={2021},
  publisher={Nature Publishing Group UK London}
}

@article{benedetti2019parameterized,
  title={Parameterized quantum circuits as machine learning models},
  author={Benedetti, Marcello and Lloyd, Erika and Sack, Stefan and Fiorentini, Mattia},
  journal={Quantum Science and Technology},
  volume={4},
  number={4},
  pages={043001},
  year={2019},
  publisher={IOP Publishing}
}

@article{Caro_2021,
	doi = {10.22331/q-2021-11-17-582},  
	year = 2021,
	optmonth = {nov},
	publisher = {Verein zur Forderung des Open Access Publizierens in den Quantenwissenschaften},
	volume = {5}, 
	pages = {582}, 
	author = {Matthias C. Caro and Elies Gil-Fuster and Johannes Jakob Meyer and Jens Eisert and Ryan Sweke},
	title = {Encoding-dependent generalization bounds for parametrized quantum circuits},
	journal = {Quantum}
}

@inproceedings{gilfuster2025relation,
title={On the Relation between Trainability and Dequantization of Variational Quantum Learning Models},
author={Elies Gil-Fuster and Casper Gyurik and Adrian Perez-Salinas and Vedran Dunjko},
booktitle={The Thirteenth International Conference on Learning Representations},
year={2025},
url={https://openreview.net/forum?id=TdqaZbQvdi}
}

@article{Schollwock_2011,
   title={The density-matrix renormalization group in the age of matrix product states},
   volume={326},
   ISSN={0003-4916},
   url={http://dx.doi.org/10.1016/j.aop.2010.09.012},
   DOI={10.1016/j.aop.2010.09.012},
   number={1},
   journal={Annals of Physics},
   publisher={Elsevier BV},
   author={Schollwöck, Ulrich},
   year={2011},
   month=jan, pages={96–192} }

@article{Bridgeman_2017,
   title={Hand-waving and interpretive dance: an introductory course on tensor networks},
   volume={50},
   ISSN={1751-8121},
   url={http://dx.doi.org/10.1088/1751-8121/aa6dc3},
   DOI={10.1088/1751-8121/aa6dc3},
   number={22},
   journal={Journal of Physics A: Mathematical and Theoretical},
   publisher={IOP Publishing},
   author={Bridgeman, Jacob C and Chubb, Christopher T},
   year={2017},
   month=may, pages={223001} }

@article{Schreiber_2023,
   title={Classical Surrogates for Quantum Learning Models},
   volume={131},
   ISSN={1079-7114},
   url={http://dx.doi.org/10.1103/PhysRevLett.131.100803},
   DOI={10.1103/physrevlett.131.100803},
   number={10},
   journal={Physical Review Letters},
   publisher={American Physical Society (APS)},
   author={Schreiber, Franz J. and Eisert, Jens and Meyer, Johannes Jakob},
   year={2023},
   month=sep }

@misc{mhiri2024constrainedvanishingexpressivityquantum,
      title={Constrained and Vanishing Expressivity of Quantum Fourier Models}, 
      author={Hela Mhiri and Leo Monbroussou and Mario Herrero-Gonzalez and Slimane Thabet and Elham Kashefi and Jonas Landman},
      year={2024},
      eprint={2403.09417},
      archivePrefix={arXiv},
      primaryClass={quant-ph},
      url={https://arxiv.org/abs/2403.09417}, 
}

@article{rahimi2007random,
  title={Random features for large-scale kernel machines},
  author={Rahimi, Ali and Recht, Benjamin},
  journal={Advances in neural information processing systems},
  volume={20},
  year={2007}
}

@misc{angrisani2025simulatingquantumcircuitsarbitrary,
      title={Simulating quantum circuits with arbitrary local noise using Pauli Propagation}, 
      author={Armando Angrisani and Antonio A. Mele and Manuel S. Rudolph and M. Cerezo and Zoe Holmes},
      year={2025},
      eprint={2501.13101},
      archivePrefix={arXiv},
      primaryClass={quant-ph},
      url={https://arxiv.org/abs/2501.13101}, 
}

@misc{lerch2024efficientquantumenhancedclassicalsimulation,
      title={Efficient quantum-enhanced classical simulation for patches of quantum landscapes}, 
      author={Sacha Lerch and Ricard Puig and Manuel S. Rudolph and Armando Angrisani and Tyson Jones and M. Cerezo and Supanut Thanasilp and Zoë Holmes},
      year={2024},
      eprint={2411.19896},
      archivePrefix={arXiv},
      primaryClass={quant-ph},
      url={https://arxiv.org/abs/2411.19896}, 
}

@misc{bermejo2024quantumconvolutionalneuralnetworks,
      title={Quantum Convolutional Neural Networks are (Effectively) Classically Simulable}, 
      author={Pablo Bermejo and Paolo Braccia and Manuel S. Rudolph and Zoë Holmes and Lukasz Cincio and M. Cerezo},
      year={2024},
      eprint={2408.12739},
      archivePrefix={arXiv},
      primaryClass={quant-ph},
      url={https://arxiv.org/abs/2408.12739}, 
}

@misc{cerezo2024doesprovableabsencebarren,
      title={Does provable absence of barren plateaus imply classical simulability? Or, why we need to rethink variational quantum computing}, 
      author={M. Cerezo and Martin Larocca and Diego García-Martín and N. L. Diaz and Paolo Braccia and Enrico Fontana and Manuel S. Rudolph and Pablo Bermejo and Aroosa Ijaz and Supanut Thanasilp and Eric R. Anschuetz and Zoë Holmes},
      year={2024},
      eprint={2312.09121},
      archivePrefix={arXiv},
      primaryClass={quant-ph},
      url={https://arxiv.org/abs/2312.09121}, 
}

@misc{shin2025newperspectivesquantumkernels,
      title={New perspectives on quantum kernels through the lens of entangled tensor kernels}, 
      author={Seongwook Shin and Ryan Sweke and Hyunseok Jeong},
      year={2025},
      eprint={2503.20683},
      archivePrefix={arXiv},
      primaryClass={quant-ph},
      url={https://arxiv.org/abs/2503.20683}, 
}

\newpage
\appendix

\end{document}